\pdfoutput=1
\RequirePackage{fix-cm}
\documentclass[twocolumn]{svjour3}
\smartqed

\usepackage{graphicx}
\usepackage{epstopdf}
\usepackage{amsmath}
\usepackage{lineno}

\begin{document}

\title{Detector monitoring with artificial neural networks at the CMS experiment at the CERN Large Hadron Collider}

\author{Adrian Alan Pol \and
        Gianluca Cerminara \and
        Cecile Germain \and
        Maurizio Pierini \and
        Agrima Seth
}

\institute{ Adrian Alan Pol 
            \at Universit\'{e} Paris-Saclay, CERN \\
            \email{adrian.pol@cern.ch}
            \and
            Gianluca Cerminara
            \at CERN
            \and
            Cecile Germain
            \at Universit\'{e} Paris-Saclay
            \and
            Maurizio Pierini
            \at CERN
            \and
            Agrima Seth
            \at CERN
}

\date{Received: date / Accepted: date}

\maketitle
\begin{abstract}

Reliable data quality monitoring is a key asset in delivering collision data suitable for physics analysis in any modern large-scale High Energy Physics experiment. This paper focuses on the use of artificial neural networks for supervised and semi-supervised problems related to the identification of anomalies in the data collected by the CMS muon detectors. We use deep neural networks to analyze LHC collision data, represented as images organized geographically. We train a classifier capable of detecting the known anomalous behaviors with unprecedented efficiency and explore the usage of convolutional autoencoders to extend anomaly detection capabilities to unforeseen failure modes. A generalization of this strategy could pave the way to the automation of the data quality assessment process for present and future high-energy physics experiments.

\end{abstract}

\section{Introduction}
\label{sec:introduction}

The Compact Muon Solenoid (CMS) experiment is a general purpose particle physics detector operating at the CERN Large Hadron Collider \cite{lhc1995large} (LHC). Data collected with the CMS detector are used in many aspects of modern particle physics, notably the discovery \cite{chatrchyan2012observation} and characterization \cite{Khachatryan:2014jba} of the Higgs boson.

The CMS detector is described in details in~\cite{chatrchyan2008cms}, together with a definition of the used coordinate system and the relevant kinematic variables. In CMS, muons are measured with detection planes instrumented with four detector technologies: drift tubes (DTs), cathode strip chambers, resistive plate chambers, and gas electron multipliers. A detailed description of the CMS muon detectors can be found in~\cite{MUO-16-001}.

Within the CMS Collaboration, physics analysis are performed on {\em good} data, selected by imposing stringent quality criteria. During data taking, a subset of the collected statistics is processed in real time, to create a set of histograms filled with a certain critical quantities. Statistical tests are performed to compare these histograms to a set of predefined reference, representing the typical detector response during normal operation conditions. Using the histogram comparison and the outcome of the tests, expert shifters acknowledge the alarms and may decide to intervene (up to stopping the data taking), depending on the evaluation of the problem severity. The knowledge of the LHC running conditions and of the history of possible issues identified in the past, are key ingredients in this decision process. Details on the infrastructure used for this Data Quality Monitoring (DQM) are given in \cite{de2015data}. The two main domains of the monitoring chain are:
\begin{itemize}
    \item {\em online monitoring}, which provides live feedback on the quality of the data while they are being acquired, allowing the operator crew to react to unforeseen issues identified by the monitoring application;
    \item {\em offline monitoring}, designed to certify the quality of the data collected and stored on disk using centralized processing (referred to as the event reconstruction, which converts detector hits into a list of detected particles, each associated with an energy and direction).
\end{itemize}
These two validation steps differ in three main aspects: 
\begin{itemize}
    \item the latency of the evaluation process; online monitoring is requested to identify anomalies in quasi real time to allow the operators to intervene promptly while the offline procedure has a typical timescale of several days,
	\item the fraction of the data which they have access to; online processing runs at a rate of $100$~Hz, corresponding to approximately $0.1\%$ of the data written to disk for analysis, while the offline processing takes as input the full set of events accepted by the trigger system ($\sim 1$~kHz of data),
	\item the granularity of the monitored detector components; while offline monitoring requires identifying only overall status of the detector components, online should determine faulty subdetector elements.
\end{itemize}
Despite their specific characteristics, these two steps rely on the same anomaly detection strategy: the scrutiny of a long list of predefined histograms, selected to detect a set of known failure modes. These histograms are monitored by detector experts, who compare each distribution to a corresponding reference, derived from {\em good data} in line with predetermined validation guidelines.
 
This two-layer monitoring protocol was adopted by the CMS Collaboration for LHC Run I (2010-2012) and in Run II (2015-2018). The ever increasing detector complexity, monitoring data volumes and the necessity to cope with different LHC running scenarios call for an increasing level of automation of the applications in the future. Already, the amount of histograms to monitor is challenging for a single shifter, while the number of histograms to monitor increases every time a new failure mode is identified and consequently added to the list of known potential problems. 
Furthermore, the human intervention and currently implemented tests require collecting a substantial amount of data, implying a detection delay. Last but not least, the cost in terms of human resources is substantial i.e. the 24/7 DQM shifter and the expert personnel responsible for updating the good data references and related instructions. We believe that introducing machine learning into the CMS DQM process will help with those challenges.

This work focuses on the online monitoring. We concentrate on the application of deep learning techniques, and specifically \textit{image-like processing}~\cite{lecun2015deep} for the automation of detector level monitoring. While the main focus of this work is on improving detection specificity and sensitivity, the proposed approach could come with practical advantages in operation being based on less astringent assumption on the nature of the anomalies.

As a concrete example we use real data recorded by the CMS DT chambers of the muon spectrometer during the data-taking campaign of the LHC Run II. 
The main aspects of this work are:
\begin{itemize}
    \item we exploit the geographical information of the detector assessing the (mis)behavior with high-granularity and then combining the results to probe different detector components;
	\item we detect different types of anomalies affecting the detector at different scales (ranging from a few channels to collective behaviors of big portion of the DT system) by combining different algorithms;
    \item we show that image-like processing achieves considerably better performance with respect to the current threshold-based DQM test (later called the {\em production algorithm}) and allows to tune the working point in terms of specificity (depending on the deployment strategy).
\end{itemize}

Although the experimental demonstration of the results presented in this paper is tied to the specificities of the DT subdetector, the procedure that we discuss have a potentially broader application scope. Mainly because the typical issues encountered with other subdetectors are analogous. This possibility is currently under investigation for other detector components of the CMS experiment.

The remainder of the paper is organized as follows. Sections~\ref{sec:cms_dt} and~\ref{sec:dataset} present in more details the problem that we want to solve and describe the utilized data set. Section~\ref{sec:ml4dqm} reviews the current state of the art in the machine learning domain of failure detection. Sections~\ref{sec:local}, \ref{sec:regional} and \ref{sec:global} present three complementary approaches to the problem. Section~\ref{sec:results} describes and discusses the results.

\section{The CMS Drift Tube muon system}
\label{sec:cms_dt}

\begin{figure}
\centering
\includegraphics[width=.4\textwidth]{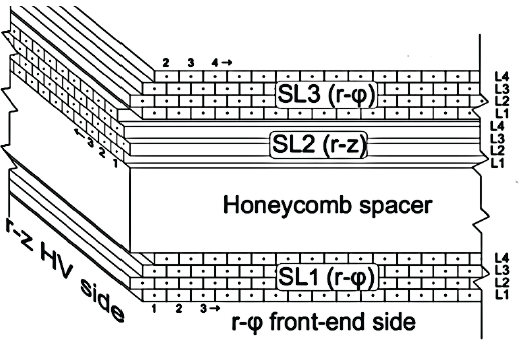}
\caption{Schematic view of the one DT chamber showing the position and orientation of the tubes. From~\cite{cms2010calibration}.}
\label{fig:dtlayout}
\end{figure}

An illustration of the internal structure of a DT chamber is shown in Fig.~\ref{fig:dtlayout}. Each chamber, on average $2\times2.5$~m in size, consists of 12 layers of drift tubes. Layers are arranged in three groups of four, each containing a variable number of tubes, up to $96$. The middle group measures the coordinate along the direction parallel to the beam and the two outer groups measure the perpendicular coordinate. Each tube corresponds to one readout channel (briefly referred to as {\em channel} in the rest of the paper). Particles carrying an electromagnetic charge and traversing a tube release an electronic signal by ionizing the gas in the tube (a {\em hit}). By combining the information provided by the channels, one can determine the trajectory of the particle crossing the chamber.

\begin{figure}
  \centering
  \includegraphics[width=.4\textwidth]{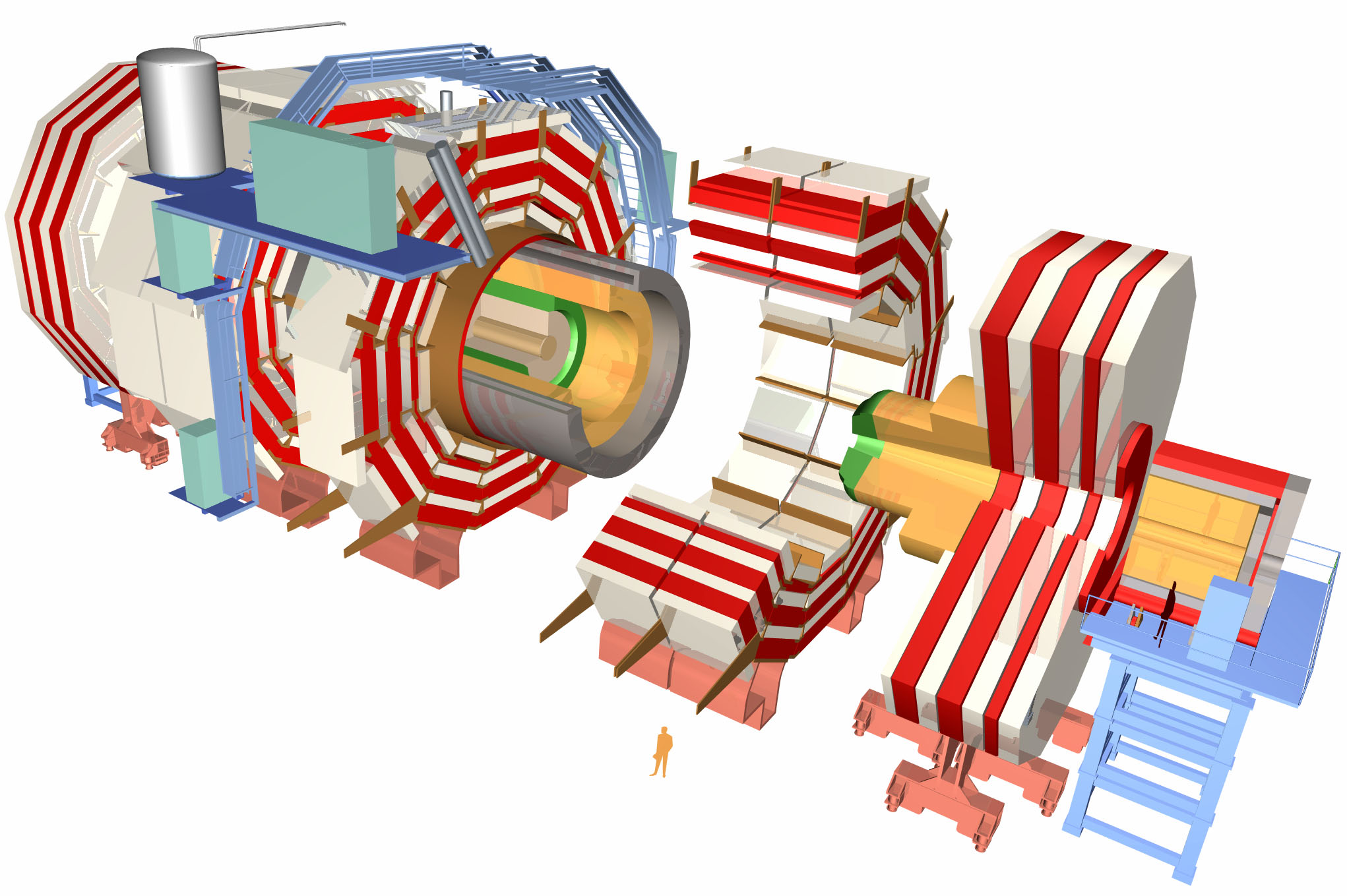}
  \caption{Magnified view of the CMS detector showing the wheel structure. The muon chambers are represented by the white volumes while the red volumes represent the iron return yoke.}
  \label{fig:cms_wheels}
\end{figure}

\begin{figure}
  \centering
  \includegraphics[width=.3\textwidth]{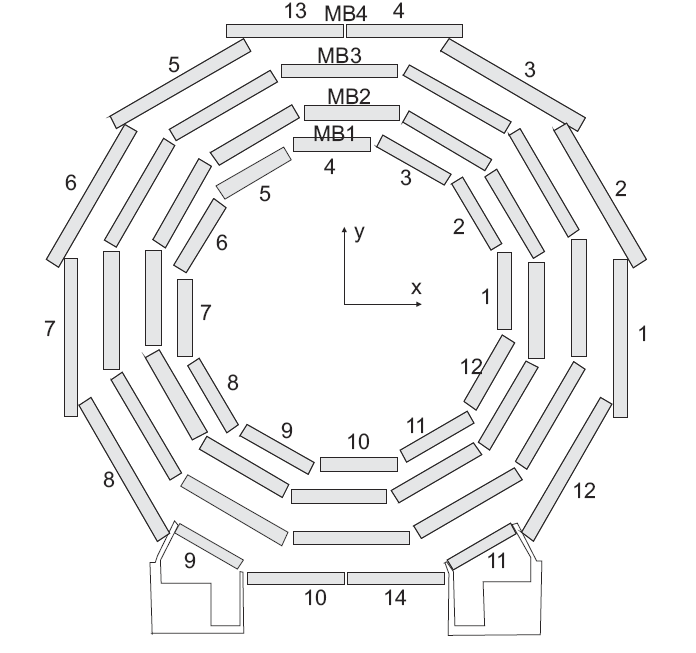}
  \caption{Numbering schema of the sector and stations of DT chambers in one wheel.  From~\cite{cms2010calibration}.}
  \label{fig:dtnumering}
\end{figure}

The chamber numbering schema follows that of the iron of the yoke, consisting in five wheels (see Fig.~\ref{fig:cms_wheels}) along the $z$-axis, each one divided into $12$ azimuthal sectors (see Fig.~\ref{fig:dtnumering}). The wheels are numbered from $-2$ to $+2$, sorted according to global CMS $z$-axis, with wheel $0$ situated in the central region around the proton-proton collision point. The sector numbering is assigned in an anti-clockwise sense when looking at the detector from the positive $z$-axis, starting from the vertically-oriented sector on the positive-$x$ side in the CMS coordinate system (sector $1$). Chambers are arranged in four stations at different radii, named MB1, MB2, MB3 and MB4. The first and the fourth stations are mounted on the inner and outer face of the yoke respectively; the remaining two are located in slots within the iron. Each station consists of $12$ chambers (one per sector) except for MB4 (which contains $14$ chambers). The total number of chambers is then $5\times(3\times12+14) = 250$.

\section{The data set, monitoring strategy, and preprocessing}
\label{sec:dataset}

\subsection{The occupancy matrix}
CMS data are organized in acquisition runs (or just {\em runs} in CMS jargon), corresponding to a given setup both of the CMS detector and of LHC accelerator. Runs are denoted by integers, increasing with time. Their duration is varying from as little as few seconds to as much as several hours.

Each run is divided into luminosity sections (LSs), a time interval corresponding to a fixed number of proton-beam orbits in the LHC and amounting to approximately $23$~seconds, numbered progressively from 1 at the start of each run. Each LS can be identified uniquely by specifying the LS number and the run number. The beam intensity (also referred to as {\em luminosity}) varies along each run, resulting in a varying number of proton-proton collision data (the events).

\begin{figure}
A \centering

\includegraphics[width=.45\textwidth]{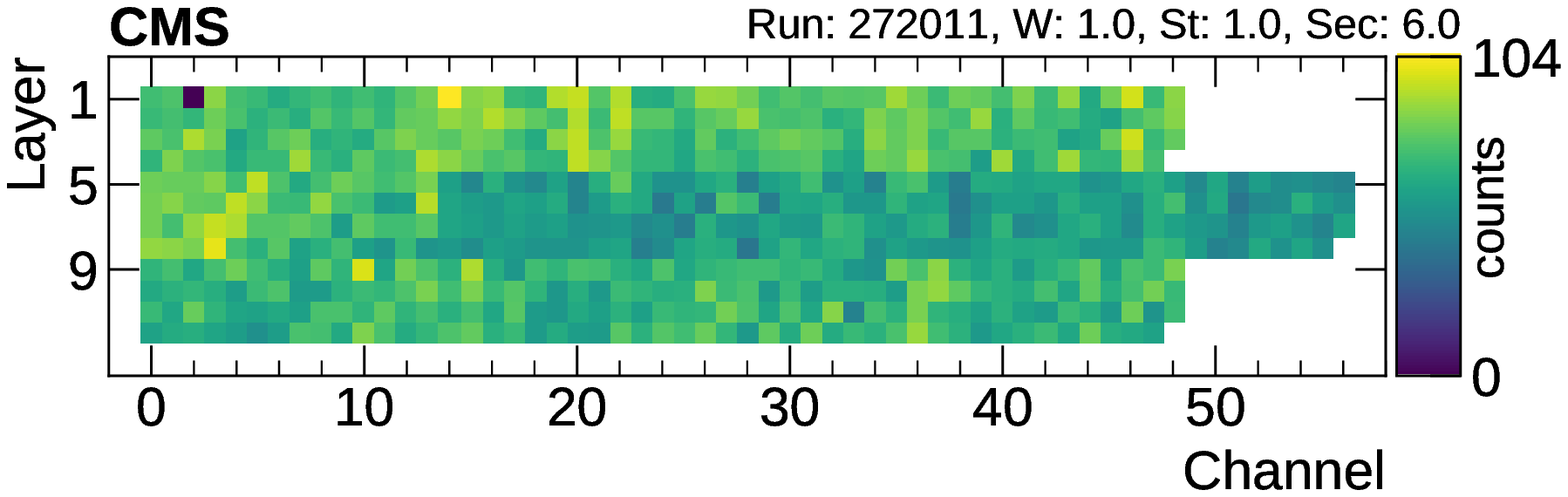}

\centering
B \centering

\includegraphics[width=.45\textwidth]{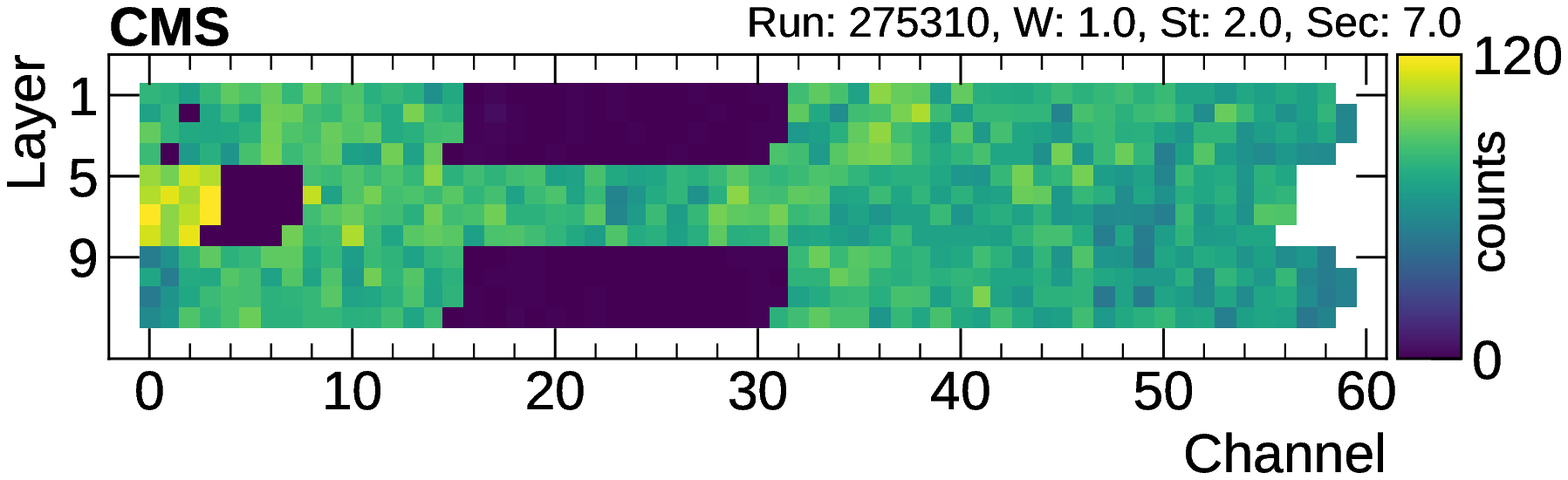}

\centering
C \centering

\includegraphics[width=.45\textwidth]{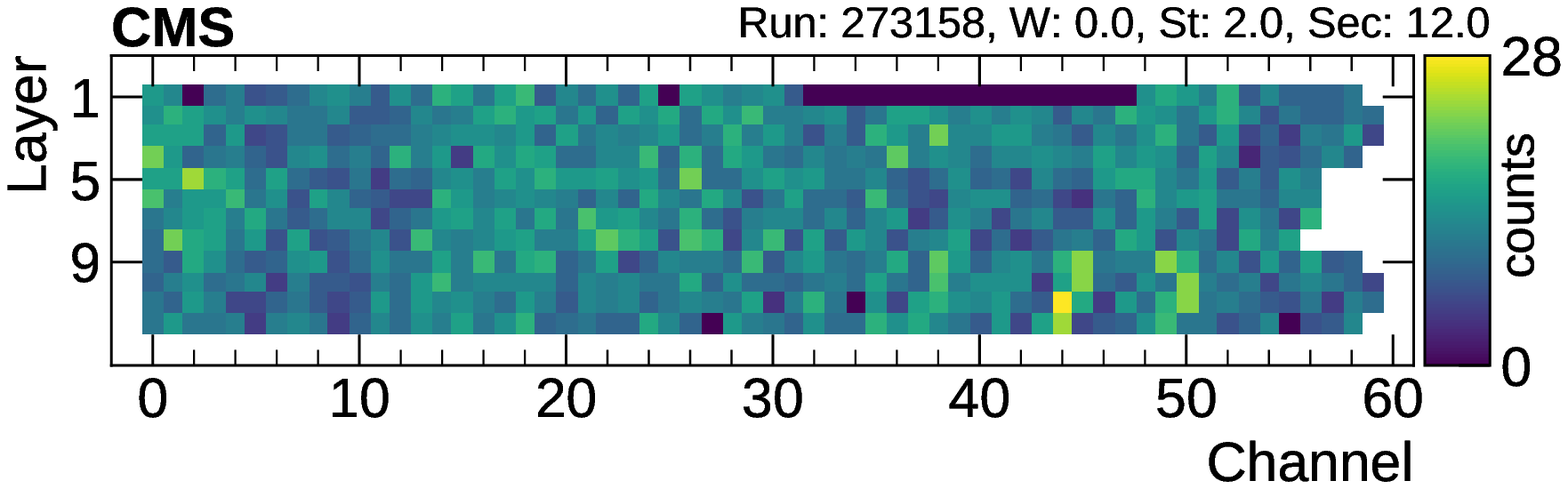}
\centering

\caption{Example of visualization of input data for three DT chambers. The data in (A) manifest the expected behavior in spite of having a dead channel in layer $1$. The {\em production algorithm} regards this instance as non-problematic. The chamber shown in the plot in (B) instead shows regions of low occupancy across the $12$ layers and should be classified as faulty. According to the run log, this effect was induced by a transient problem with the detector electronic. (C) suffers from a region in layer $1$ with lower efficiency, which should be identified as anomalous. The {\em production algorithm} classify the chamber in (B) as anomalous. However it is not sensitive enough to flag the chamber in (C).}
\label{fig:1}
\end{figure}

For each chamber $k$ in a given run, the current DQM infrastructure \cite{tuura2010cms} records an {\em occupancy matrix} $C^{k}$, which contains the total number of particle hits at each channel for a given LS or set of consecutive LSs. The occupancy matrix can be viewed as a varying size two-dimensional array organized along layer (row) and channel (column) indices:
$$ C^{k} = \{x_{i,j}^{k}; 1 \leq i \leq l, 0 \leq j < n_i\},$$
where $l=12$ is the number of layers and $n_i$ is the number of channels in layer $i$. In general, we label the chambers and their components as $C^{k}$ and $x_{i,j}^{k}$. For simplicity, we omit the $k$ index when discussing problems related to individual chambers, until Section~\ref{sec:regional}. Figure~\ref{fig:1} shows examples of occupancy matrices, represented as two-dimensional occupancy plots. The utilized data set consists of 21000 occupancy matrices for the 250 chambers.

\subsection {Monitoring strategy}
The anomaly detection method currently used in the online monitoring production system targets a specific failure scenario: a region of cells not providing any electronic signal, large enough to affect the track reconstruction in the chamber. This is by far the most frequent issue, usually related to transient problems in the readout electronics. Examples of this kind of failures are shown in Fig.~\ref{fig:1}~B~and~C. These kinds of occupancy plots are created accumulating data in  time. Once in a while, the plot filling process is reset, to increase sensitivity to problems occurring during the run. The production algorithm evaluates samples per chamber. Although it quantifies the fault severity on the basis of the fraction of affected channels, it does not identify specific faulty layers.

The novel approach proposed in this work goes beyond the functionalities of the current production algorithm. Starting from the identification of layers with under-performing cells, it provides effective identification of faulty chambers. Moreover, it exploits the geographical information of the layer and chamber position to identify different kind of failures. To this purpose, three complementary approaches are considered:
 
\begin{itemize}
    \item {\em Local}: data collected in each layer are treated independently from the others. As for the production algorithm, this approach regards chambers which have occupancy of hits with small variance between neighboring channels as expected behavior and targets a well known list of problems with a supervised approach. Chambers which have dead, inefficient or noisy regions, are considered problematic. We explore this approach in Section~\ref{sec:local}.
    \item {\em Regional}: we extend the local approach to account for intra-chamber problems, to be applied whenever faults are spotted only when the information about all layers within one chamber is present. For this purpose, we simultaneously consider all layers in a chamber, but each chamber is considered independently from the others, Section~\ref{sec:regional}.
    \item {\em Global}: we simultaneously use the information of all the chambers for a given run. The position of the chamber in the CMS detector (uniquely determined by the wheel, station, and sector numbers) impacts expected occupancy distribution of the channel hits. This approach is described in Section~\ref{sec:global}.
\end{itemize}

\subsection{Preprocessing}
\label{sec:preprocessing}
\begin{figure}
A \centering

\includegraphics[width=.45\textwidth]{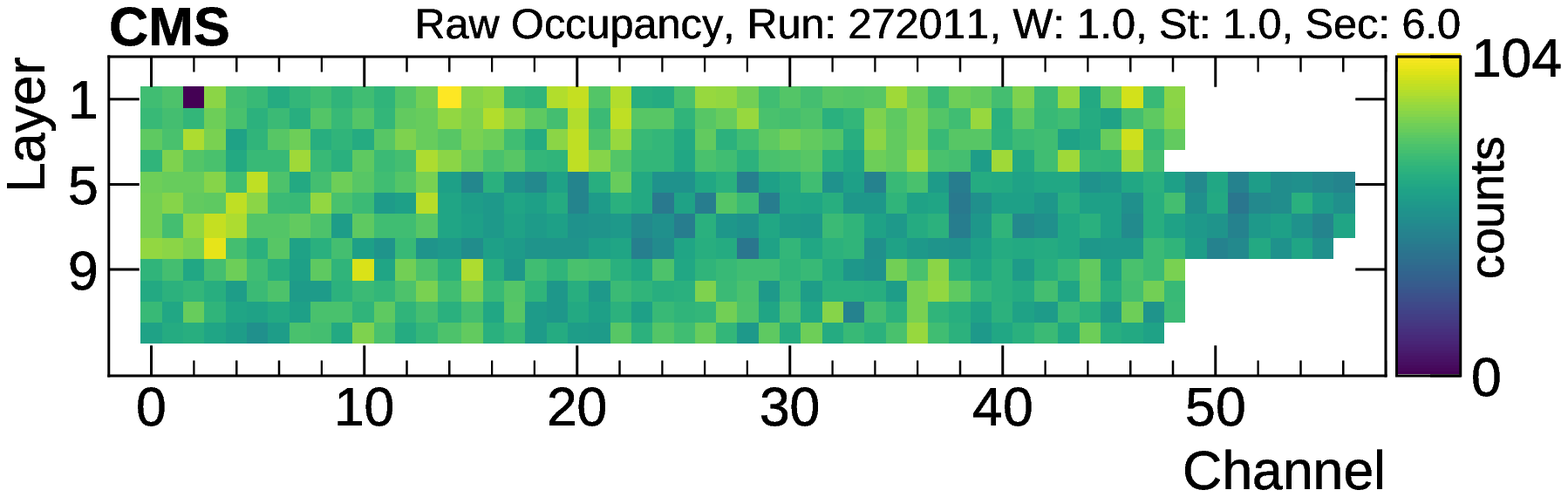}

B \centering

\includegraphics[width=.45\textwidth]{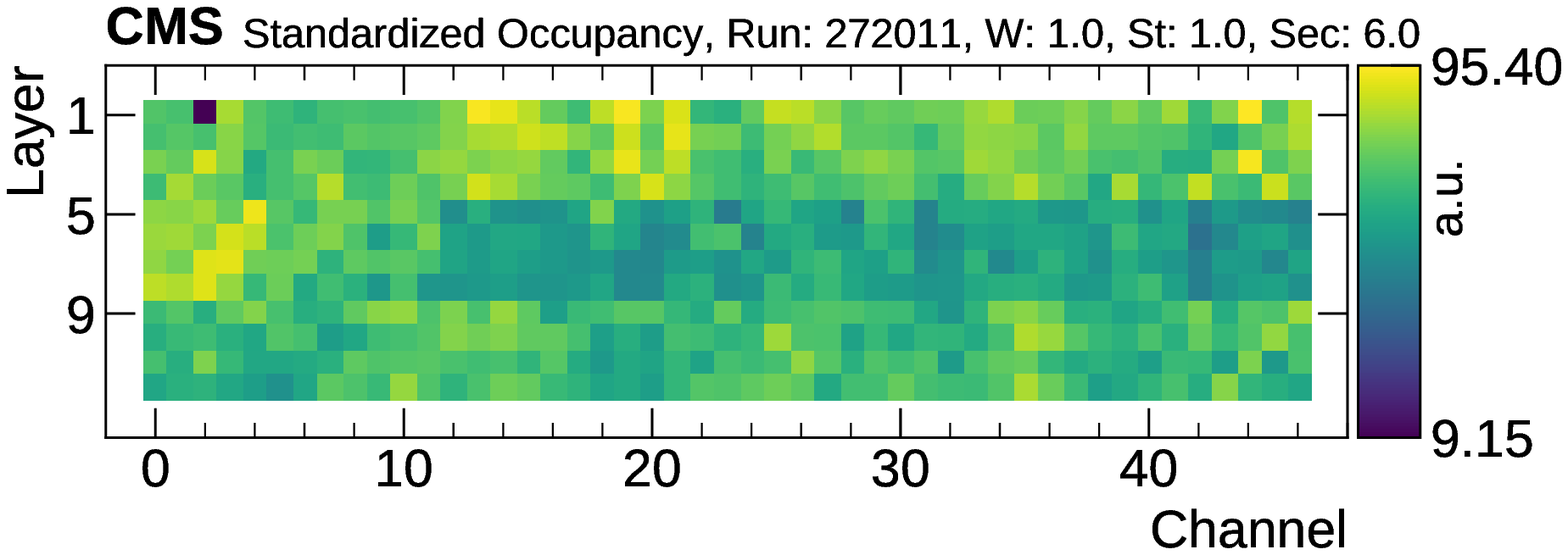}

C \centering

\includegraphics[width=.45\textwidth]{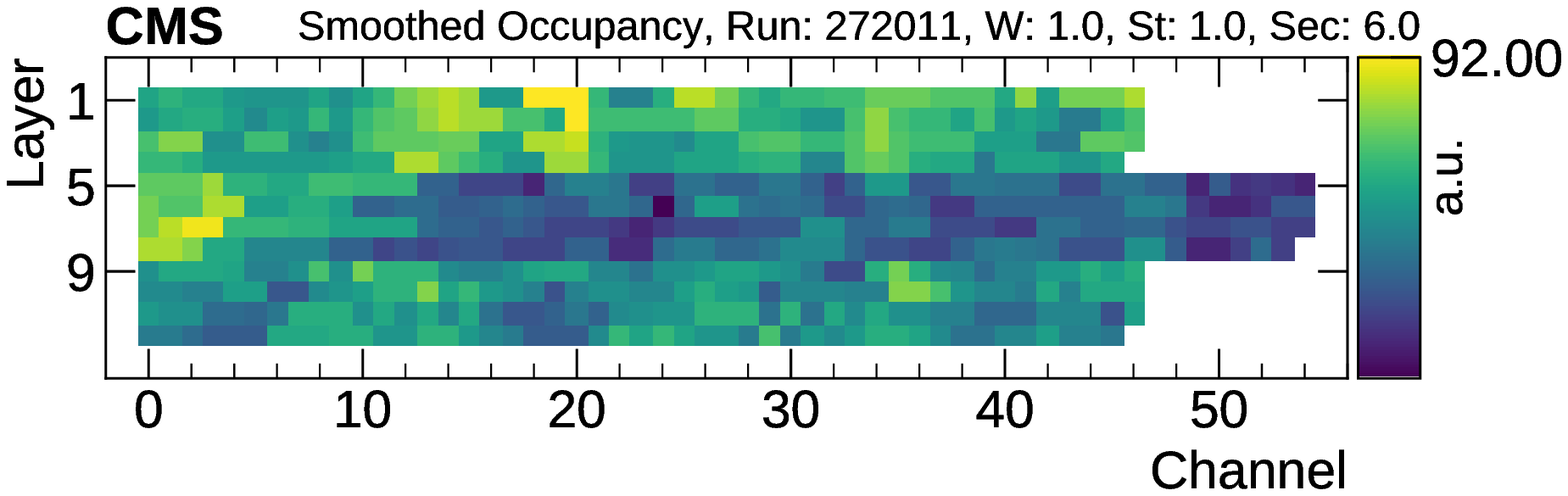}

D \centering

\includegraphics[width=.45\textwidth]{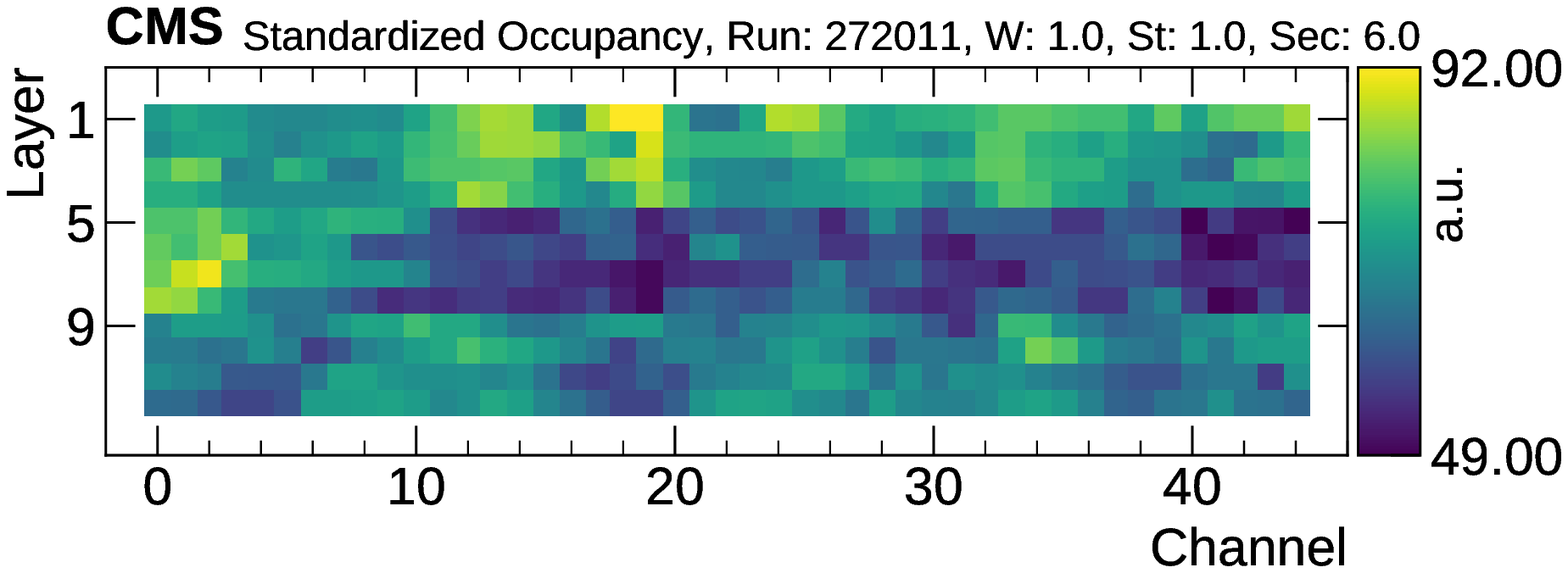}
\centering

\caption{Example of two kinds of input sample preprocessing. (A) acquired ({\em raw}) values, (B) standardizing each layer directly from raw values using linear interpolation. (C) smoothing the raw values data with median filter (D) standardizing each layer from smoothed data. In (C), the isolated low-occupancy spot in layer 1, corresponding to a dead channel, is discarded.}
\label{fig:2}
\end{figure}

A common data set preprocessing procedure is used for the three studies (for visual interpretation, see Fig.~\ref{fig:2}).

\begin{itemize}
  \item {\em Standardization of the chamber data}: the number of channels $x$ in a layer varies not only within the chamber but also depends on the chamber position in the detector. This quantity falls between $47$ and $96$. We enforce a fixed-input dimensionality by applying a layer-by-layer one dimensional linear interpolation to match the size of the smallest layer $s$ in data set. The smallest layer is chosen to simplify our models later in this study. Starting from the recorded matrix $x_{ij}$, a standardized matrix $\tilde{x}_{i, j}$ is defined as:
$$\tilde{x}_{i, j} = frac(\alpha) (x_{i, \lfloor \alpha \rfloor} - x_{i, \lceil \alpha\rceil}) + x_{i, \lfloor \alpha \rfloor}.$$
where $\alpha$ is an interpolation point, defined by $\alpha = j \frac{n_{i}}{n_{s}}$. We verified that this method doesn't compromise sensitivity to very small problematic regions despite a small reduction in the amplitude and sharpness of the anomalies.
  \item {\em Smoothing}: according to CMS DT experts, misbehaving channels are problematic only when a spatially contiguous cluster of them is observed. Instead, isolated misbehaving channels are not considered a problem. To take this into account the one dimensional median filter is applied:
$$\hat{x}_{i, j} = \text{med} (x_{i, j}, x_{i, j+1}, x_{i, j+2}).$$
\item {\em Normalization}: the occupancy of the chambers in the input data set depends on the integration time and on the LHC beam configuration and intensity i.e. on the number of LSs spanned when creating the image and corresponding luminosity. The normalization strategy depends on the need for comparing data across chambers or across runs: the precise procedure used in the two approaches is described in Sections~\ref{sec:local} and \ref{sec:regional}, respectively.
\end{itemize}

\section{Machine learning for DQM anomaly detection}
\label{sec:ml4dqm}

In this section we briefly discuss machine learning anomaly detection techniques in light of both the operational condition and the a priori knowledge of the data. Machine learning presents several advantages over the currently adopted procedure as the decision function can be learned from collected data. In the future, it might be possible to bypass human intervention when the algorithm decision is not controversial and only invoke the shifters opinion for intermediate questionable cases. An example of this approach is discussed in \cite{borisyak2017towards} in the context of the CMS offline monitoring. The high data dimensionality precludes simple parametric density estimation of the normal behavior. This leaves an extremely wide range of methods such as one-class Support Vector Machine ($\mu$-SVM) \cite{scholkopf2001estimating}, Isolation Forest \cite{liu2008isolation,liu2012isolation} and different flavors of deep learning. For a general survey see~\cite{aggarwal2015outlier}.

Anomaly detection techniques usually assume rarity of abnormal events (considered as outliers with respect to the normal generating process) and/or lack of a complete set of representative examples of all possible behaviors. If such representative examples are available, anomaly detection reduces to binary classification (supervised learning), with possibly the help of various resampling methods \cite{aggarwal2014data} or reformulation of the objective function \cite{cowan2011asymptotic} for dealing with class imbalance.

In our case, supervised learning is clearly a valid option as specific anomalous scenarios were extensively studied. The CMS DQM framework keeps copious archives of subdetector-specific quality-related quantities, e.g. the DT occupancy plots. Moreover, the imbalance between good and bad data is not extreme, with a typical rate of anomalies reaching the 10\% level. These anomalies are then frequent enough for a sizable set of them to be used for supervised training.

However, there is a good motivation for a semi-supervised anomaly detection approach. Beside the deep learning methods that will be discussed at the end of this section, we experiment with the two reference methods, which are variants of one-class classification: $\mu$-SVM and Isolation Forest. $\mu$-SVM estimates the support of the data distribution by a non-linear (kernel) transform of the data space (as in all SVM techniques) and by identifying the hyperplane that maximizes the separation of the training data from the origin. Accordingly, $\mu$-SVM has the important property of being a novelty detection algorithm: once trained, it is not sensitive to the frequency of anomalies. However, the implicit prior of kernel-based classification is that the function to be learned is smooth such that generalization can be achieved by local interpolation between neighboring training examples. As argued at length by Bengio et al. (for instance in \cite{bengio2007scaling} and \cite{bengio2013representation}), this assumption is questionable for high data dimensionality. An alternative is the Isolation Forest, which copes with the curse of dimensionality by relying only on the principle of isolation of outliers in a random recursive partitioning of the feature space along the axes and tree ensembles. The Isolation Forest algorithm does not rely on any distance or density measure, but assumes that anomalies can be isolated in the native feature space. Besides being highly scalable to large data sets, Isolation Forest offers some possibility of interpretation. 

Classical fully unsupervised approaches based on neighborhood (e.g. k nearest neighbors), topological density estimation (Local Outlier Factor and its variants) or clustering (for a detailed presentation, see \cite{goldstein2016comparative}) are not relevant here. These algorithms have quadratic complexity and poorly perform in high dimensions, because of data sparsity (in high dimensions, all pairs of points become almost equidistant)~\cite{zimek2012survey}. Moreover, a simple geometric (e.g. Euclidean) distance in the feature space does not define a similarity metric. For instance, the distance between examples A and B in Fig.~\ref{fig:1} is dominated by 
the contribution of well-behaving channels. The similarity function, or equivalently the adequate representation, must be learned from the data. 

This representation learning view \cite{bengio2013representation} points towards deep learning, as it should remain sensitive to the local geometric relationship in the data related to the underlying apparatus. Convolutional networks \cite{krizhevsky2012imagenet} integrate the basic knowledge of merely the topological structure of the input dimensions and learn the optimal filters that minimize the objective error.

A more ambitious goal is to extract an explanatory representation of the anomalies with latent variables, in a probabilistic framework (e.g. restricted Boltzmann machines, or variational autoencoders~\cite{Rezende:2014}), where the learned representation is the posterior distribution of the latent variables given an observed input. However, even the inference step with these representations may suffer from high computational cost, and requires some further feature construction.

The alternative is the trade-off between interpretability and simplicity by learning a direct encoding, typically as a neural network based autoencoder, which is a parametric map from inputs to their representation. Although it has been argued that, even for basic neural networks, most of the training is devoted to learning a compressed representation~\cite{Tishby:2015,Tishby:2017}, autoencoders are particularly suitable for anomaly detection. When trained on the inliers, testing on unseen faulty sample tend to yield sub-optimal representations, indicating that a sample is likely generated by a different process. In order to go beyond simple dimensionality reduction while preventing over-fitting, various flavors of regularization are proposed (the literature being considerable, we give only some entry points): 
\begin{itemize}
\item sparse autoencoders~\cite{Ranzato:2006} penalize the output of the hidden unit activations or the bias;
\item denoising autoencoders~\cite{Vincent:2010} robustify the mapping by requiring it to be insensitive to small random perturbations;
\item contractive autoencoders~\cite{rifai_contractive_2011} pursue the same goal, by penalizing analytically the sensitivity of learned features in a data-driven interpretation of the Tangent Propagation algorithm~\cite{simard_transformation_1998}.
\end{itemize}
In fact, denoising and contractive autoencoders learn density models implicitly, through the estimation of statistics or through a generative procedure~\cite{Alain:2014}.

\section{Local approach: detecting faulty behavior within a layer}
\label{sec:local}

\subsection{Motivation}

The first experiment concentrates on training a classifier to identify local problems, i.e. considering each layer independently from the others. This approach enforce the expert knowledge of what is currently considered correct or anomalous and probes the detector with higher granularity than the production algorithm. The goal is to identify regions of channels not registering any hits (called {\em dead channels} in detector jargon), or having lower detection efficiency (hence lower hit counts with respect to the neighboring ones in the same layer) or being dominated by electronic noise (called {\em noisy channels}). These are by far the most frequent failure modes. The local approach can be considered as an initial benchmark comparing fully supervised, semi-supervised and unsupervised methods, and specific algorithms in each category, before embarking in full-fledged anomaly detection. Moreover, the local approach, if successful, can be further exploited as a preprocessing step for filtering these trivial faults before attempting to detect more elusive ones.

Given the locality restriction of this approach, contextual information is not accessible. As a consequence of this, a model based on this strategy will not be able to spot, for example, a faulty layer in which occupancy is decreased uniformly with respect to neighboring layers. We acknowledge this limitation and address it in Section~\ref{sec:regional}.

\subsection {Data set and methods}
After having applied the standardization procedure (see Section~\ref{sec:preprocessing}), a layer is represented as a single row of an occupancy matrix: 
$$X_{i}= (\tilde{x}_{i, 1}, \tilde{x}_{i,2}, \ldots ,\tilde{x}_{i,47}).$$
The available data set consists of 21000 chambers corresponding to 228480 individual layers.

Hit counts in a layer are normalized to a $[0,1]$ range, dividing them by the maximum of the occupancy value in the layer:
$$ \dot{x}_{i,j}=\frac{\tilde{x}_{i,j}}{\max(X_{i})}.$$
The need for normalization comes from the intrinsic variation of the occupancy, which depends on the spatial position of the chamber (as described in more detail in Section~\ref{sec:global}) and on the integration time of the analyzed image. 

In this experiment, we compare the performances of the following:
\begin{itemize}
\item unsupervised learning with (a) a simple statistical indicator, the variance within the layer, and (b) an image processing technique, namely the maximum value of the vector obtained by applying a variant of an edge detection Sobel filter \cite{sobel1990isotropic}:
$S_i=\max(\begin{bmatrix} -1 & 0 & 1 \end{bmatrix} * X_i)$.
\item semi-supervised learning, with (c) Isolation Forest, and (d) $\mu$-SVM.
\item supervised learning, with (e) a fully connected shallow neural network (SNN), and (f) a convolutional neural network (CNN); 
\end{itemize}

The ground truth is established by field experts on a random subset of the data set, by visually inspecting the input sample before any preprocessing: 5668 layers were labeled as good and 612 as bad. The 9,75\% fault rate is a faithful representation of the real problem at hand. With this ratio, both anomaly and outlier detection approach can be considered. Out of this set 1134 good and 123 bad examples are reserved to compose the test set corresponding to 20\% of the labeled layers. The remaining examples are used for training and validation for the semi-supervised and supervised methods.

\begin{figure}
\includegraphics[width=.467\textwidth]{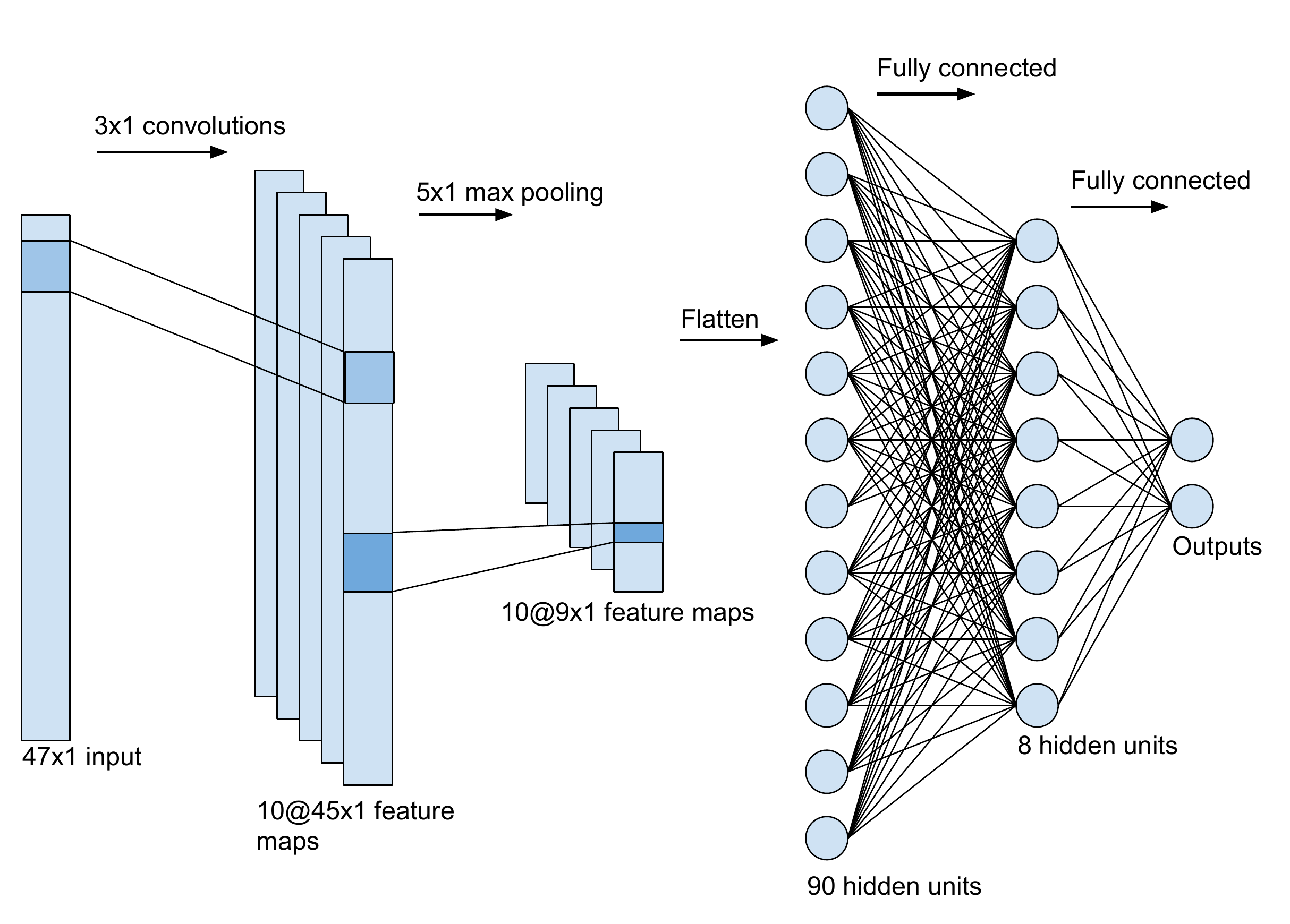}
\centering
\caption{Architecture of the convolutional neural network model used to target the local strategy.}
\label{fig:3}
\end{figure}

The Isolation Forest and $\mu$-SVM models are cross-validated using five stratified data set folds to search for their corresponding optimal hyper-parameters. Subsequently, the Isolation Forest is retrained using those hyper-parameters ($100$ base estimators in the ensemble) on the full unlabeled data set, while $\mu$-SVM (RBF kernel, $\nu$ of $0.4$, $\gamma$ of $0.1$) is retrained using only negative class examples. The architecture of the CNN model with one dimensional convolution layers used for this problem is shown in Fig.~\ref{fig:3}. Rectified linear units are chosen as activation functions for inner-layer nodes, while the softmax function is used for the output nodes. The model is trained using the Adam \cite{kingma2014adam} optimizer and early stopping mechanism monitoring validation set (set to 20\% of data set) with patience set to 32 epochs. The model is implemented in Keras \cite{chollet2015keras}, using TensorFlow \cite{abadi2016tensorflow} as a backend.

The SNN model consists of one hidden fully-connected layer with 16 units (chosen to approximately match number of parameters in the CNN). As for CNN, it uses rectified linear unit as activation function of the hidden nodes and the softmax function is used for the output nodes. This model is primarily introduced to obtain a term of comparison for the CNN. 

Unlike what was done for the other models, we do not apply the smoothing preprocessing step described in Section~\ref{sec:preprocessing} for CNN nor SNN models, in order to allow them to learn their filters. Additionally, we weight our negative $S_{0}$ and positive $S_{1}$ samples to account for class imbalance. The weight $\lambda_{\psi}$ for a sample in class $\psi \in \{0,1\}$ is defined by
$$ \lambda_{\psi} = \frac{| S |}{2 \cdot | S_{\psi} |} $$
where $S = S_{0} \cup S_{1}$. We discuss the results in Section~\ref{sec:loc_res}.

\section{Regional approach: detecting unusual behavior within a chamber}
\label{sec:regional}

\subsection {Motivation}

In normal conditions, healthy chambers show similar occupancy levels in neighboring layers. The four central chambers have a different behavior due to their different orientation (see Section~\ref{sec:cms_dt}). The regional approach exploits the relative occupancy patterns of the layers within a chamber. For example, it aims at detecting failure modes where the occupancy of hits decreases uniformly in a specific layer or set of layers. Typical examples of these kind of failures are problems related with the high-voltage bias of the drift cells. The voltage distribution system is organized by layers and a lower value w.r.t to the nominal operation point would result in lower detector efficiency and, as a consequence, lower absolute occupancy in the affected region. Figure~\ref{fig:voltage-fault} shows an example of such an occurrence, where layer 9 is misbehaving. The production algorithm and the local models of Section~\ref{sec:local} are not conceived to detect this type of anomalies.

\subsection {Data set and methods}

In the early stages of this work we observed that a model capable of detecting regional anomalies cannot be successfully trained if the local faults are not filtered beforehand. Moreover, the available $\sim 500$ labeled images do not provide a sufficiently large training set. Thus we start with a much larger data set (all the unlabeled samples). We solve the labeling problem using the score of the convolutional model presented in Section~\ref{sec:local} as an approximation of the ground truth. For this, we choose a working point with 99\% true positive rate (to guarantee a large data set size) and 5\% false positive rate. Approximately $90\%$ of the collected data are labeled as good by online and offline monitoring experts. We then estimate the residual bad example contamination to be $\sim 0.5\%$. We believe that the residual contamination of problematic chambers is reduced to a tolerable level. All chambers with any layer identified as faulty are discarded. Chambers located in MB4 are discarded as well, because of the lack of a middle group of four layers, see Section~\ref{sec:cms_dt}. The above changes effectively narrowed the training data set to 8441 matrices. The smoothing and standardization procedures are applied to all the layers $\tilde{C}^{k}$ within each chamber obtaining matrices of 
shape $12 \times 46$. 
The occupancy of hits within one chamber are normalized using a min-max scaler:
$$\dot{C}= \frac{\tilde{C}^{k} - \min(\tilde{C}^{k})}{\max(\tilde{C}^{k}) - \min(\tilde{C}^{k})}.$$
This normalized values to the $[0, 1]$ range while retaining the information about the relative occupancy between the layers.

\begin{figure}
\centering A
\centering
\includegraphics[width=.467\textwidth]{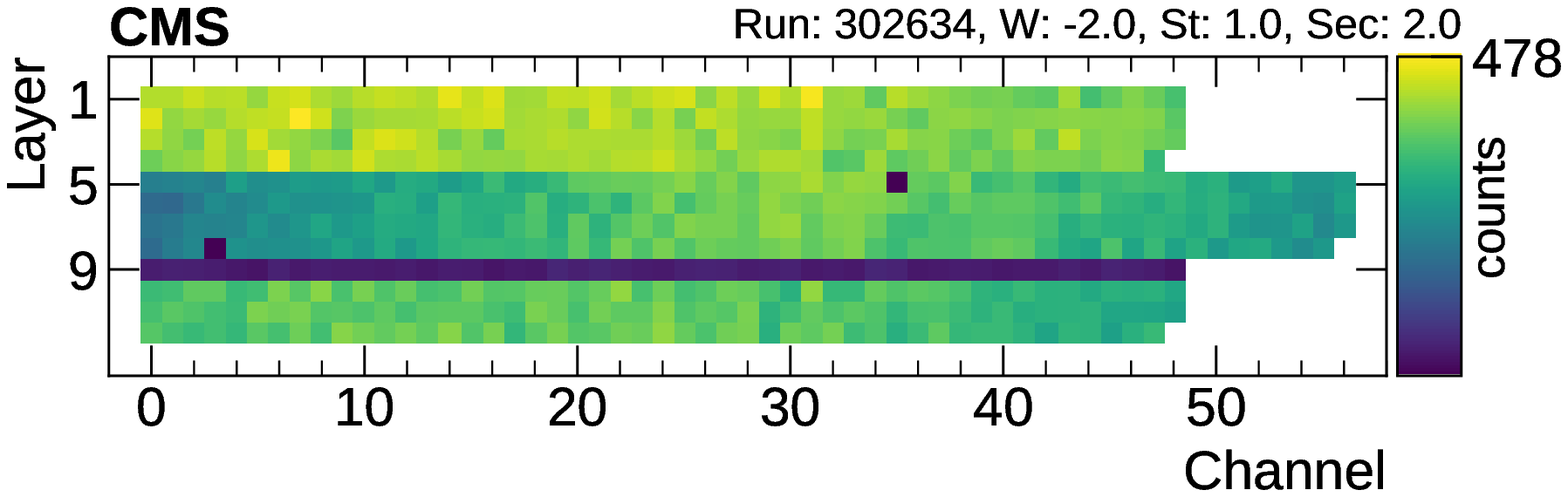}
\centering B
\centering
\includegraphics[width=.467\textwidth]{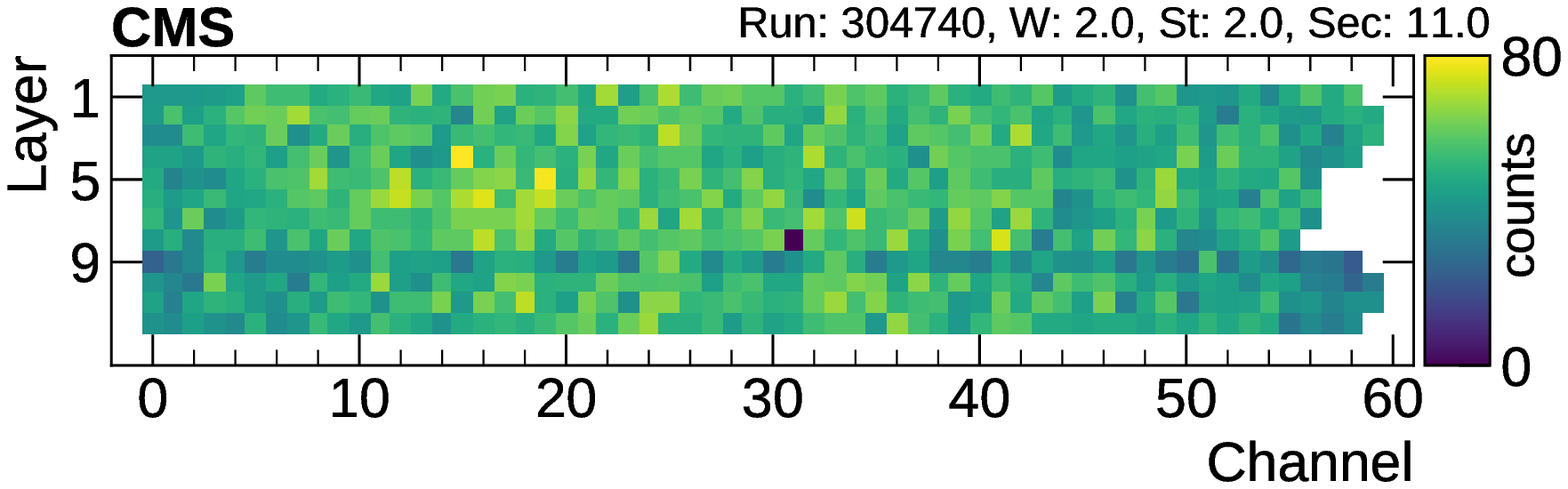}
\centering

\caption{Example of the impact on hit counting of different voltages applied to layer 9. (A) shows the occupancy map when operating the layer at $3200$~V and (B) shows the effect of operating at $3450$~V. Both examples should be regarded as anomalies. Since the values in both cases are not equal to zero, the {\em production algorithm} considers those cases as non-problematic.}
\label{fig:voltage-fault}
\end{figure}

In order to evaluate the model, we use the only labeled set for the class of anomalies that we want to tackle: a subset of the data (runs 302634 and 304737-304740), during which layer 9 of some chambers was operating at a voltage lower than the nominal one (see Fig.~\ref{fig:voltage-fault}). In particular, the voltage was set to $3450$~V in runs 304737-304740 and to $3200$~V in run 302634 while the standard operation point is $3550$ or $3600$~V depending on the chamber.
These settings result in an absolute difference in hit counting, more pronounced for the lower voltage settings, because the physics of gas ionization by radiation. 
The chambers where all layers operate at nominal conditions are considered as {\em good} in the test.

\begin{figure}
\centering
A

\centering
\includegraphics[width=.467\textwidth]{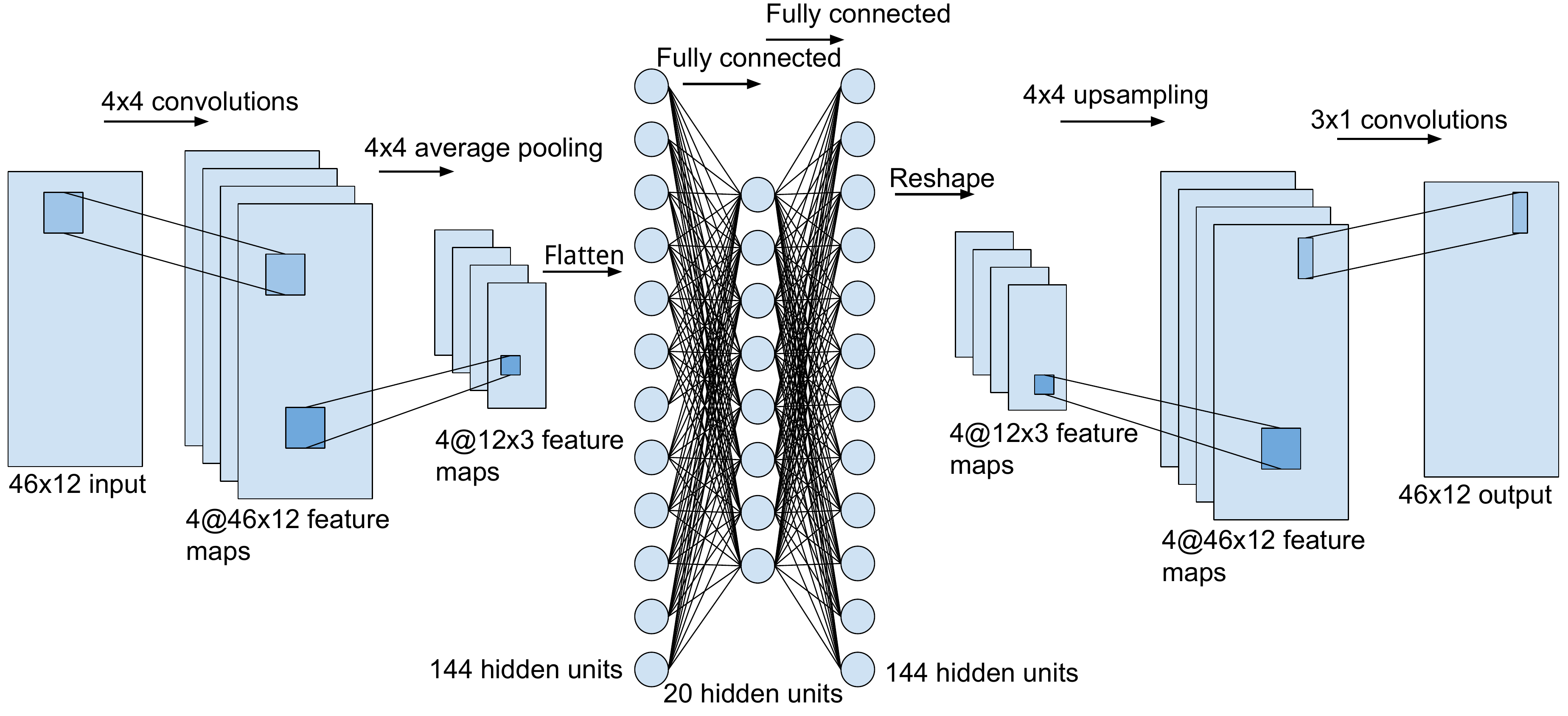}

\centering
B

\centering
\includegraphics[width=.467\textwidth]{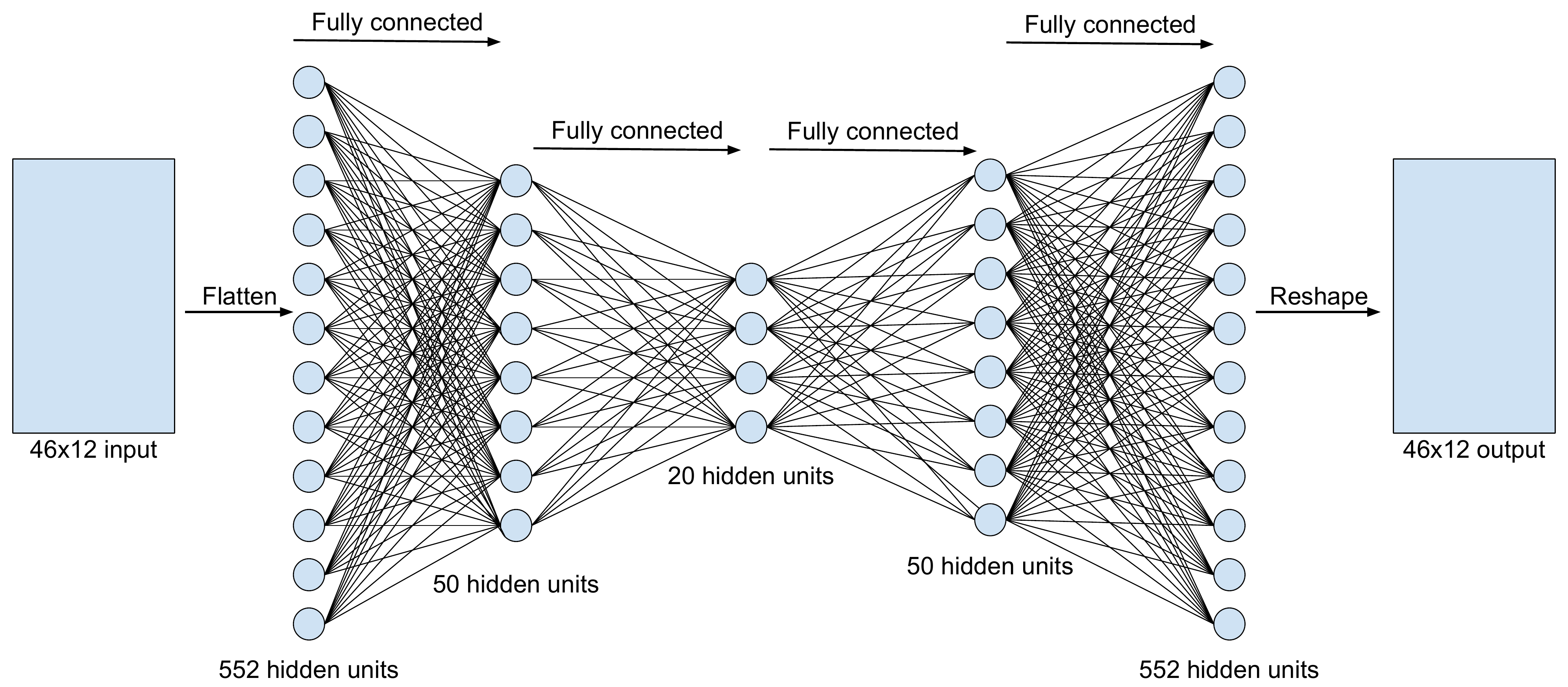}
\caption{Convolutional (A) and simple, denoising, sparse (B) autoencoder models architecture used to target {\em regional} strategy.}
\label{fig:convae}
\end{figure}

In this experiment, the following semi-supervised approaches are considered:
\begin{itemize}
	\item simple bottleneck autoencoder with the representation layer equal to $20$ units;
	\item convolutional autoencoder;
	\item denoising autoencoder in which we add additional artificial noise to training samples;
	\item autoencoder with kernel L1 ($10^{-5}$) sparsity regularization in the hidden layers.
\end{itemize}
Similarly to the local approach we train the autoencoders using the Adam optimizer. Early stopping mechanism with the patience set to 32 epochs is adopted to monitor validation set ($20\%$ of the total data set). 
All models are implemented using the Keras library with TensorFlow as a backend.
The architecture of the model is shown in Fig.~\ref{fig:convae}. A and B for, respectively, the convolutional autoencoder and the other three models (for which a common architecture is adopted). The bottleneck architecture is kept for both denoising and sparse autoencoders in order to limit the amount of parameters to train. The parametric rectified linear unit is used as the activation function on the hidden layers, while the output layer uses the sigmoid function. All models are instructed to minimize the mean squared error (MSE) $\epsilon$ between original, $\dot{x}$, and reconstructed, $\ddot{x}$, samples: 
$$\epsilon^{k} = \frac{1}{ij}\sum_{i,j} (\dot{x}^{k}_{i,j} - \ddot{x}^{k}_{i,j})^{2}.$$
We discuss the results in Section~\ref{sec:reg_res}.

\section{Global approach: detecting unusual behavior using global information}
\label{sec:global}

\subsection {Motivation}
In the third approach, we aim at detecting anomalies looking at the global ensemble of muon chambers, exploiting the dependency of the occupancy of each of them on their position in the detector. 
We categorize the chambers according to their position in the spectrometer and its impact on the occupancy pattern, exploiting the field knowledge to predetermine the classes of chambers.

The expected occupancy pattern is mainly driven by the proximity to the beam-collision point, at the center of the detector. As a consequence, chambers in different stations (see Section \ref{sec:cms_dt}) will manifest a different behavior. The rotational symmetry of the detector geometry and of the collision events around the beam axis is taken into account grouping chambers within the same station, independently on the sector they belong to (see Section \ref{sec:cms_dt}). Similarly chambers belonging to the same station but in opposite wheels are considered alike. 
Additionally, the behavior of the chambers is expected to be the same across different runs, modulo the overall decrease of occupancy due to the decrease of beam intensity across the fill.
This leaves us with a categorization based on the chamber numbering schema, where the station number and the absolute value of the wheel number are the only relevant parameters.

The problem is clearly contextual, in the sense that important explanatory attributes are not part of the basic data features. Conditional anomaly detection \cite{song2007conditional} has been proposed to deal with such a situation when the relevance of external attributes is unknown. For instance, if a set of environmental or technical attributes are monitored that can impact the behavior of the detector components. In our case, the spatial position of the chambers are our only external attribute, and their impact is assured by common understanding of the underlying physics processes. Thus, we are back to a point anomaly problem.

\subsection {Methods}
In this approach we use a bottleneck autoencoder similar to that introduced in Section~\ref{sec:regional} (see Fig.~\ref{fig:convae}), except that the size of the bottleneck layer is reduced to three units for visualization purposes. We also follow the same preprocessing, training and validation procedure.
The goal of the study is to exploit the categorization of the chambers based on their geographical location to interpret the compressed representation of the network.

Global faults are not tracked before by DT experts. In absence of a global label, we only considered an unsupervised method for this experiment. We discuss the results in Section~\ref{sec:glo_res}.

\section{Results and Discussion}
\label{sec:results}

\subsection{{\em Local} approach}
\label{sec:loc_res}

\begin{figure}
\includegraphics[width=.46\textwidth]{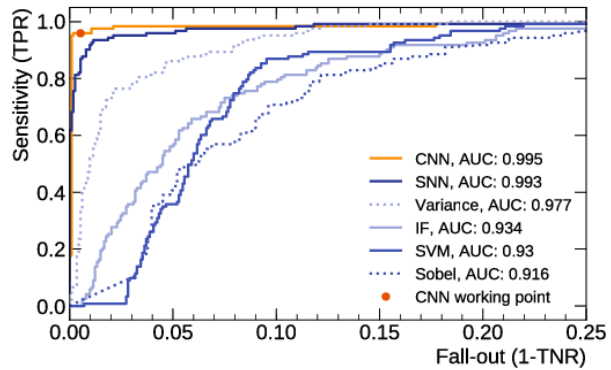}
\centering
\caption{ROC curves for different models used in the \textit{local approach. The Area Under the Curve (AUC) is quoted to compare the performance.}}
\label{fig:5}
\end{figure}

\begin{figure}
\includegraphics[width=.46\textwidth]{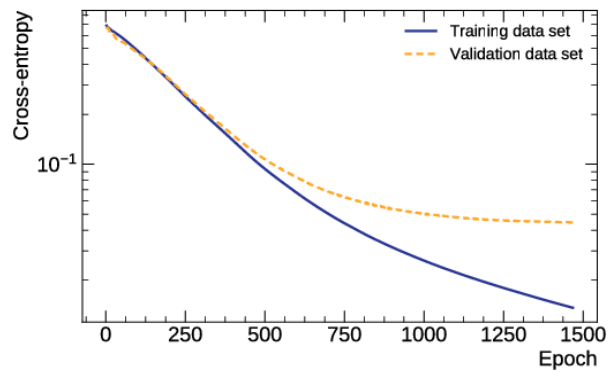}
\centering
\caption{Loss function as a function of the number of epochs in the training of the CNN model used for the local approach. The two curves illustrate the behavor of the training and validation data sets.}
\label{fig:epoch}
\end{figure}

The performance of the various models on a held out test data set can be seen in Fig.~\ref{fig:5}, where we show the different receiver operating characteristic (ROC) curve. Compared to statistical, image processing or other machine learning based solutions, supervised deep learning clearly outperforms the rest. Thanks to the limited number of parameters of the model, the training converges to a satisfactory result (Fig.~\ref{fig:epoch}), despite the number of training samples being small.

Although the Area Under Curve (AUC) of the fully-connected shallow neural network is comparable to the one of CNN, the latter is a better solution when requiring maximum specificity (true negative rate (TNR), aims at avoiding false positives) and sensitivity (true positive rate (TPR), aims at avoiding false negatives).
The relatively good performance of the basic and unsupervised variance method, compared to the poor results of the filter, and the near optimal performance of the SNN, show that the features to learn are not simple contrasts, although the superior performance of the CNN demonstrate that the initial edge detection layer is useful. The limited performance of Isolation Forest is likely to come from the violation of its fundamental assumption, that faults are rare (remember that the fault rate is in the order of 10\%) and homogeneous. The inferior performance of the typical semi-supervised method ($\mu$-SVM) illustrates the well-known smoothness versus locality argument for deep learning \cite{bengio2007scaling,bengio2013representation}: the difficulty to model the highly varying decision surfaces produced by complex dependencies involving many factors.

\begin{figure}
\includegraphics[width=.46\textwidth]{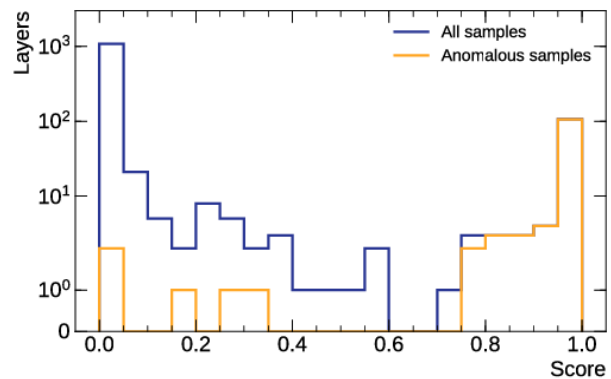}
\centering
\caption{Distribution of scores in \textit{local approach} for the CNN model.}
\label{fig:6}
\end{figure}

As shown in the score distribution of Fig.~\ref{fig:6}, the proposed architecture of the CNN model separates anomalous layers significantly. This allows for great flexibility in choosing the working point for deployment in production in the CMS DQM. Depending on the cost of type 1 and type 2 errors for the detector operators 
the threshold can be set anywhere in $[0.1, 0.9]$ score range. When using the CNN for the selection of good samples for training the regional algorithms, the working point is chosen not to favor specificity nor sensitivity, with a threshold equal to score 0.5.

The production algorithm targets a specific failure scenario of dead regions and produces a chamber-wise goodness assessment, without being capable of identifying a specific problematic layer in the chamber. For this reason we can not directly compare its performance with our local approach. For the sake of benchmarking our approach, we use our per-layer ground truth to label as bad any chamber with at least one problematic layer. We then ask the production algorithm if it indicates there is at least one faulty layer in a chamber. With this per-chamber label, we are able to estimate the specificity of the production algorithm to 91\%, with a sensitivity of only 26\%.

\begin{figure}
\includegraphics[width=.45\textwidth]{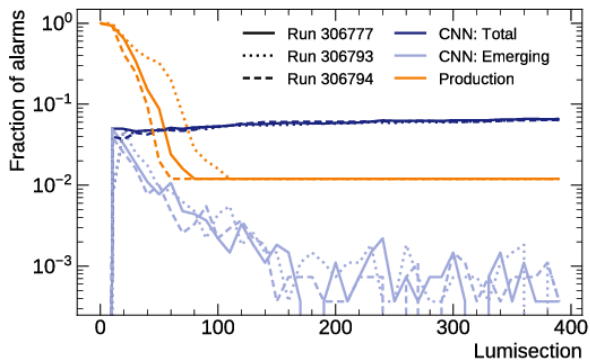}
\centering

\caption{Stability of the proposed model and the {\em production algorithm} as a function of time ({\em number of lumisections}) for three different runs: 306777, 306793, 306794. The stability test is performed every $10$ LSs. The \textit{CNN: Total} and \textit{Production} lines follow the total fraction of alarms generated by the methods. Instead \textit{CNN: Emerging} reports the fraction of new alarms being generated by the CNN model w.r.t previous test point.}
\label{fig:stability}
\end{figure}

Another difference of our approach with respect to the production one is the performance with low statistics i.e. at the beginning of a run. As seen in Fig.~\ref{fig:stability}, our CNN model gradually adds alarms until reaching stability. The production algorithm has the opposite behavior, generating a substantial fraction of false alarms in the early stages of the run. 

\subsection{{\em Regional} approach}
\label{sec:reg_res}

To assess the performance of a given ensemble of channels we take as anomaly indicator the quantity: 
$$\epsilon^{k}_{i} = \frac{1}{j}\sum_{j} (\dot{x}^{k}_{i,j} - \ddot{x}^{k}_{i,j})^{2}~,$$
i.e., the MSE between the original sample given as input to the encoder ($\dot{x}^{k}_{i,j}$) and the output of the decoder ($\ddot{x}^{k}_{i,j}$). With the objective of identifying the problematic region of the chamber, we exploit the granularity of the autoencoder information computing the MSE values for different set of channels. For example, we can compute the MSE for all the channels corresponding to a given read-out electronic board or, alternatively we can compute it per layer when tackling potential failures of the voltage distribution system.

\begin{figure}
\includegraphics[width=.467\textwidth]{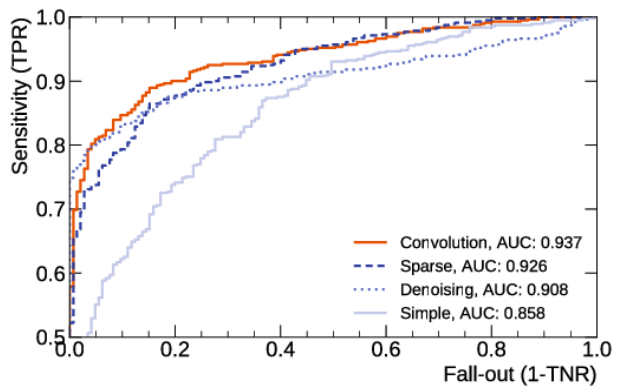}
\centering
\caption{ROC and AUC of the different autoencoder models used in \textit{regional approach}. The discriminator between good and anomalous samples is the $\epsilon$ in layer 9.}
\label{fig:aeerror}
\end{figure}

\begin{figure}
\centering
A

\includegraphics[width=.467\textwidth]{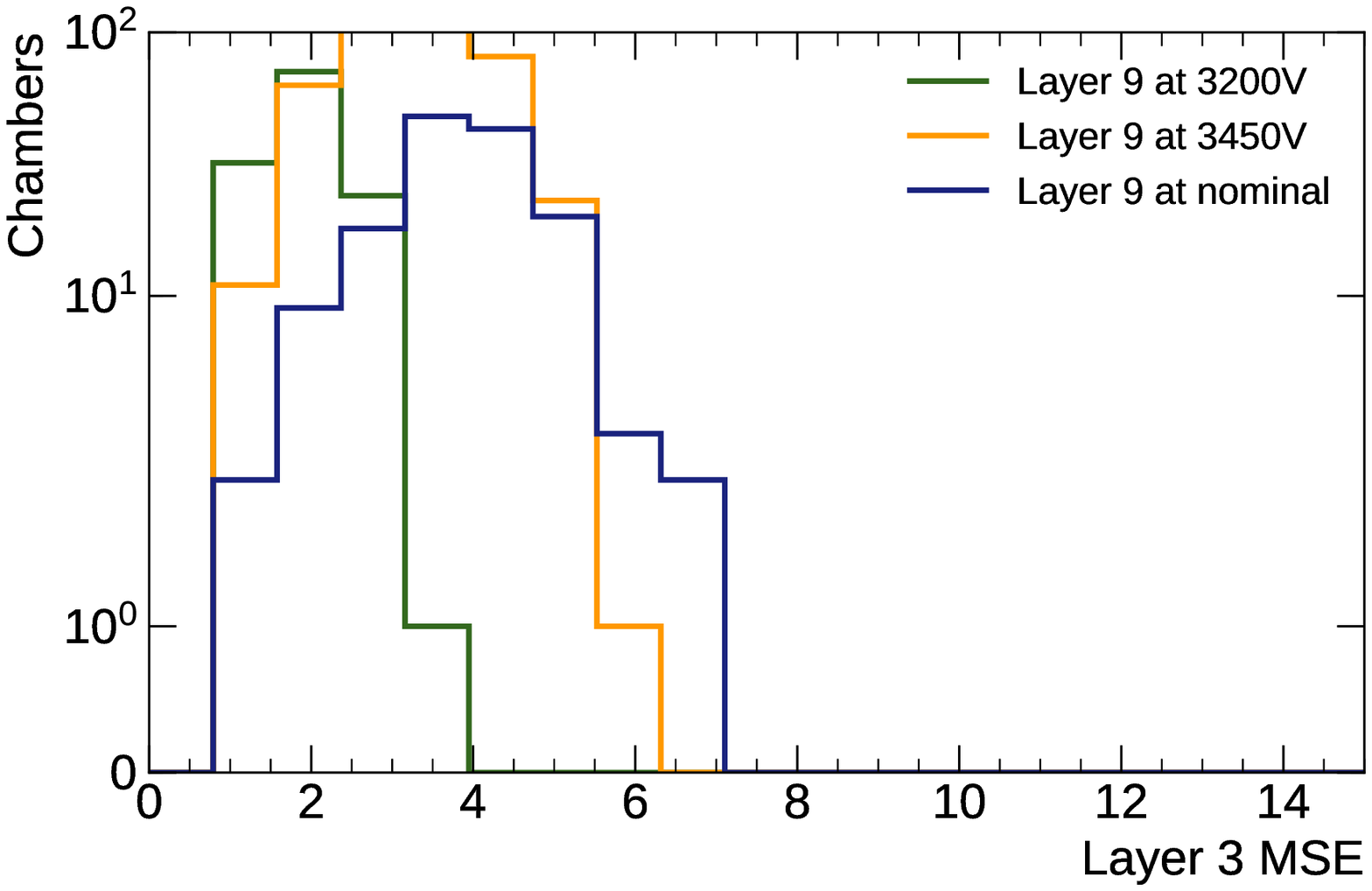}
\centering
B

\includegraphics[width=.467\textwidth]{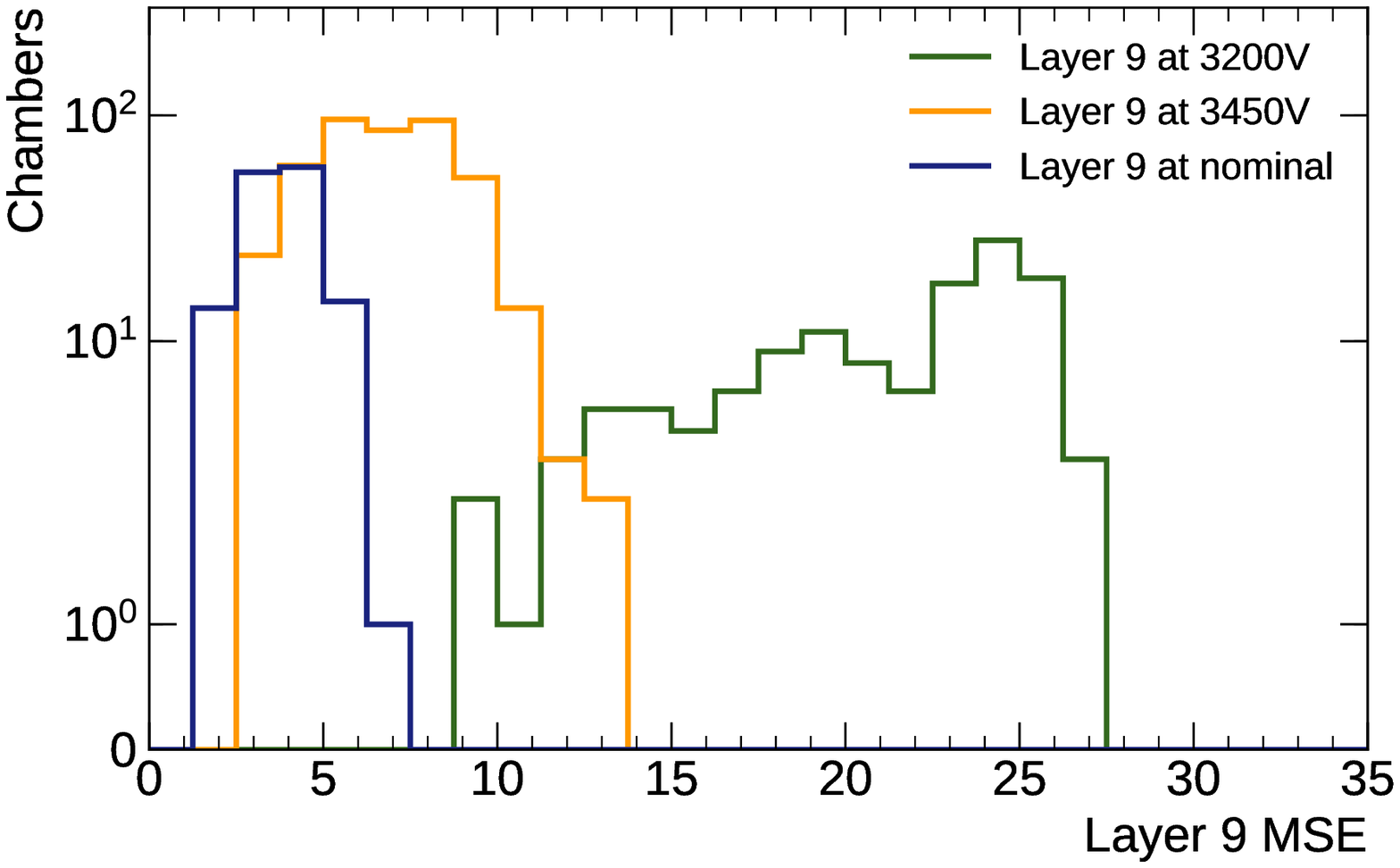}
\centering

\caption{MSE between reconstructed and input samples for layer 3 (A) and layer 9 (B) for 3 categories of data for convolutional autoencoder. Despite a problem in layer 9, all $\epsilon$ for layer 3 are comparable for all chambers.}
\label{fig:aeerror2}
\end{figure}

We use this figure of merit on the sample with different voltage settings described in Section~\ref{sec:regional}. Figure~\ref{fig:aeerror} shows good performance of all models, especially convolutional autoencoder. The distributions of the MSE for a well behaving and a problematic layer are shown in Fig.~\ref{fig:aeerror2}. The MSE distribution for layer 9 shows clear separation for chambers operated at nominal and lower voltages. For each $\epsilon_{i}$ value for a given example, a quantitative assessment of the severity of a potential anomaly can be derived quoting the corresponding p-value of the good example distribution. The separation is less pronounced for the working point at $3450$~V being closer to the nominal setup. This reflects in the AUC values reported in Fig.~\ref{fig:aeerror}.

The production algorithm is not sensitive to the type of faults described in this section since the hits in layer 9 are non-zero values. Thinking about deployment in the DQM infrastructure of the CMS experiment, the best result would be obtained when applying the local and regional models in a pipeline.

\begin{figure}
\includegraphics[width=.467\textwidth]{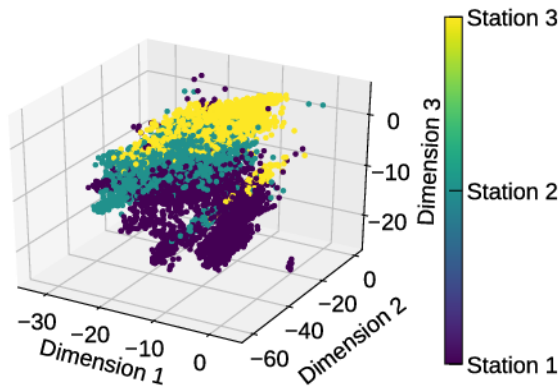}
\centering
\caption{Compressed representation of the chamber-level data of the {\em global} model. The samples cluster according to position in the detector. Here depending on the station number.}
\label{fig:latentA}
\end{figure}

\begin{figure}
\includegraphics[width=.467\textwidth]{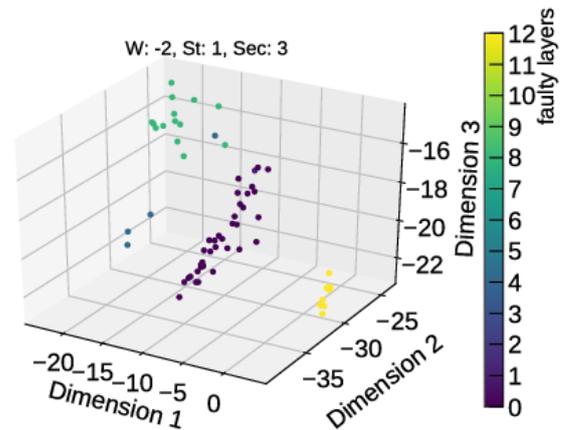}
\centering
\caption{Compressed representation of the chamber-level data of the {\em global} model limited to only one chamber across different runs with respect to number of faulty layers (scale). The samples cluster according to similar behavior.}
\label{fig:latentB}
\end{figure}

\subsection{{\em Global} approach}
\label{sec:glo_res}

Figure~\ref{fig:latentA} shows an example of a low-dimensionality representation of the chamber data clustering depending on the chamber position in the detector. 
The global approach is then potentially capable to spot an unusual behavior of DT chambers taking into account the geographical constraints. Ultimately, this could pave the way to more flexible assessment by scoring per detector region.

When investigating the representations for a specific chamber across different runs (see Fig.~\ref{fig:latentB}), we notice that the representations tend to cluster depending on the number of problematic layers. Thanks to this fact, the cumulative distribution of the compressed representation could be used to highlight the occurrence of new anomalies or to associate an anomalous behavior to an already known problem. This application could assist experts in diagnosing transient and reoccurring issues.

\section{Conclusions and Outlook}

This paper shows how detector malfunctions can be identified with high accuracy by a set of automatic procedures, based on machine learning. We considered the specific case of the DT muon chambers of the CMS experiment. We developed a CNN-based classifier to spot local misbehaviors of the kind currently targeted by the existing CMS monitoring tools.  We also showed that it is possible to extract more information from the map of electronic hits than the currently implemented statistical tests. In particular, we developed a strategy to spot regional problems across layers in a detector chamber, or globally, i.e., across chambers in the full muon detector. These algorithms, based on autoencoders,  will offer a more robust anomaly detection strategy, not being defined as supervised classifiers of specific failure modes. This approach allows to localize the origin of a given anomaly, exploiting the granularity offered by the use of MSE of the decoded image as a quantification of the anomaly. 

Currently, these algorithms have been integrated into the CMS online DQM infrastructure and they are being commissioned with the early data of the 2018 Run. The model could be further refined, e.g. integrating a mechanism of periodic retraining that would allow to repeat alarms for known problems, or to adapt to the long-term changes of the detector and beam conditions.

Since CNN is the basic ingredient in this study, and since many monitored quantities in typical high-energy physics experiments are based on 2D maps (e.g., detector occupancy, detector synchronization, etc.), the approach proposed in this paper could be extended beyond the presented use case. We hope that this case study could serve as a concrete showcase and could motivate further DQM automation using machine learning. 

\section*{Acknowledgements}

We thank the CMS collaboration for providing the data set used in this study. We are thankful to the members of the CMS Physics Performance and Data set project and the CMS DT Detector Performance Group for useful discussions, suggestions, and support. We acknowledge the support of the CMS CERN group for providing the computing resources to train our models and of CERN OpenLab for sponsoring A.S.'s internship at CERN, as part of the CERN OpenLab Summer student program. We thank Danilo Rezende for precious discussions and suggestions. This project has received funding from the European Research Council (ERC) under the European Union's Horizon 2020 research and innovation program (grant agreement n$^o$ 772369). 

\bibliographystyle{unsrt}
\bibliography{csbc}
\end{document}